\documentclass[12pt]{iopart}

\usepackage{graphicx}
\begin{document}

\def\be{\begin{equation}}
\def\ee{\end{equation}}

\def\bc{\begin{center}}
\def\ec{\end{center}}
\def\bea{\begin{eqnarray}}
\def\eea{\end{eqnarray}}
\newcommand{\avg}[1]{\langle{#1}\rangle}
\newcommand{\Avg}[1]{\left\langle{#1}\right\rangle}
\newcommand{\cor}[1]{\textcolor{red}{#1}}

\title[Percolation on interacting, antagonistic networks]{Percolation on interacting, antagonistic networks }

\author{Kun Zhao}

\address{  Department of Physics, Northeastern University, Boston,
Massachusetts 02115,USA}

\author{Ginestra Bianconi}

\address{ School of Mathematical Sciences, Queen Mary University of London, London E1 4NS, United Kingdom}
\ead{ginestra.bianconi@gmail.com}

\begin{abstract}
Recently,  new results on percolation of interdependent networks have shown that  the percolation transition can be first order.
In this paper we show that, when considering antagonistic interactions between interacting networks, the percolation process might present a bistability of the equilibrium solution.
To this end, we introduce   antagonistic interactions  for which the functionality, or activity, of a node in a network is incompatible with the functionality, of the linked nodes in  the  other interacting networks.
In particular, we study the percolation transition in two interacting networks with purely antagonistic interaction and different topology.
\end{abstract}


\maketitle

\section{Introduction}
\label{intro}
Over the last ten years  percolation processes,  and more in general,   dynamical   processes in complex networks  \cite{crit,Dyn}, have gathered  great attention.
In this context it has been shown that complex topologies strongly affect the dynamics occurring in  networks. However, many complex systems involve interdependencies between different networks, and accounting for these interactions is crucial in economic markets, interrelated technological and  infrastructure systems, social networks, diseases dynamics, and human physiology.  Recently, important new advances have been made in the characterization of percolation  \cite{Havlin1,Vespignani,Grassberger,Dorogovtsev,Havlin3,Havlin2,Gao,Havlin_dep} and other dynamical processes  \cite{Yamir1,Yamir2,Jesus,Ivanov,Goh,Goh2} on interacting and interdependent networks.
In these systems, one network function depends on the operational level of other networks. A failure in one network could  trigger failure avalanches in the other interdependent network, resulting in the increased fragility of the interdependent system.
In fact, it has been  shown {\cite{Havlin1,Vespignani,Grassberger,Dorogovtsev,Havlin3}} that two interdependent networks are more fragile than a single network and that the percolation transitions in interdependent networks can be first order.
These results have subsequently been  extended to multiple interdependent networks \cite{Havlin2,Gao} and to networks in which only a fraction of the nodes are interdependent \cite{Havlin_dep}.

Here, we want to investigate the role of antagonistic interactions in the percolation transition between interacting networks.
For antagonistic interactions  the functionality, or activity, of a node in a network is incompatible with the functionality, of the linked nodes in the other interacting networks.
As it is happening in spin systems, where antiferromagnetic interactions can result in  the frustration of the system, also in interacting network, the presence of antagonistic interactions between the nodes introduce further complexity in the percolation problem.
As a first step in investigating this complexity in this paper we will consider  two interacting networks with purely antagonistic interactions. Moreover we assume,  that a node is active in a network only if it belongs to the giant component of active nodes in the network.
We will show that for two Poisson networks with exclusively antagonistic interactions the  percolating configuration   corresponds to the percolation of one of the two networks.
Nevertheless, the solution of the model is  surprising because there is a wide region of the phase space in which there is a bistability of the percolation process:  either one of the two networks might end up to be percolating. Therefore,  in this new percolation problem, not only the percolation transitions might be    first order, but we found that there is
a real hysteresis in the system as we modify  the average degrees of the two networks.
Furthermore, we extend the analysis  to networks with other topologies, studying the percolation transition in two antagonistic scale-free networks, and in two networks  one of which is a Poisson network, and the other   one is a scale-free network.
We characterize the  rich phase digram of the percolation transition in these networks.
Interestingly, in the percolation phase diagram of these interacting networks there is a region in which both networks percolate, demonstrating a strong interplay between the  percolation process and the topology of the network.
Finally, these results shed new light on the complexity that the percolation process acquires, when considering percolation on interdependent, antagonistic networks.

\section{Percolation on antagonistic networks}
\label{sec:1}
%and \cite{RefJ}
%\subsection{Subsection title}
%\label{sec:2}
%as required. Don't forget to give each section
%and subsection a unique label (see Sect.~\ref{sec:1}).
%\paragraph{Paragraph headings} Use paragraph headings as needed.

In this paper we introduce antagonistic interactions in percolation  on interdependent networks. As it has been done for the studied case of interdependent networks \cite{Havlin1,Havlin_dep,Grassberger} we will assume that a node is active in a network only if it belongs to the giant component of active nodes in that network. The difference with respect to the case of interdependent networks is that  if a node $i$ is active on one network it cannot be active in the other one.
We consider two networks of $N$ nodes. We call the networks, network A and network B  with   degree distribution $p^A(k),p^{B}(k)$ respectively. Each node $i$ is represented in both networks.
In particular, each node has a set of neighbor nodes $j$ in network A, i.e. $j\in N^A(i)$ and a set of neighbor nodes $j$ in network B, i.e.  $j\in N^B(i)$.

A node $i$ belongs to the percolation cluster of   network A,  if it  has at least one neighbor $j\in N^A(i)$  in the percolating cluster  of  network A,  and has no neighbors  $j\in N^B(i)$  in network B  that belong to the percolating cluster of network B.
Similarly, A node $i$ belongs to the percolation cluster of   network B,  if it  has at least one neighbor $j\in N^B(i)$  in the percolating cluster  of  network B,  and has no neighbors  $j\in N^A(i)$  in network A  that belong to the percolating cluster of network A.
The percolation steady state can be found by a message passing algorithm \cite{Mezard,Grassberger} (called  by Son et al.  epidemic spreading).
Each node $i$ sends a message to each of his neigboring nodes $j$. We call each message $y^{A(B)}_{i\to j}$ if the message is sent from a node $i$ to a node $j$ in the network $A(B)$. The message $y^{A(B)}_{i\to j}$ indicates the probability that following a link $(i,j)$ in network $A(B)$ from $j$ to $i$ we reach a node $i$ which is active in  the network A(B).
The probability $S_i^{A (B)}$ that a node $i$ is active in network A (network B) depends on the messages $y^{A(B)}_{k\to i}$ that the neighbors $k$ on network A and network B send to node $i$, i.e.
\begin{eqnarray}
S^A_i&=&\left[1-\prod_{k\in N_A(i)}(1-y^A_{k \to i})\right]\prod_{k\in N_B(i)} (1-y^B_{k\to i})\nonumber\\
S_i^B&=&\left[1-\prod_{k\in N_B(i)}(1-y^B_{k \to i})\right]\prod_{k\in N_A(i)} (1-y^A_{k\to i}).
\label{ant_S}
\end{eqnarray}
Moreover the messages $y^{A(B)}_{i\to j}$ on a locally tree-like network are the fixed point solution  $n\to \infty$ of the following iterative equations for $y^{A(B),n}_{i\to j}$
\begin{eqnarray}
y^{A,n}_{i\to j}&=&\left[1-\prod_{k\in N_A(i) \setminus j}(1-y^{A,n-1}_{k \to i})\right]\prod_{k\in N_B(i)} (1-y^{B,n-1}_{k\to i})\nonumber \\
y^{B,n}_{i\to j}&=&\left[1-\prod_{k\in N_B(i) \setminus j}(1-y^{B,n-1}_{k \to i})\right]\prod_{k\in N_A(i)} (1-y^{A,n-1}_{k\to i})
\label{ant_message_rec}
\end{eqnarray}
In order to find the messages, usually the variables  $y^{A(B),n}_{i\to j}$ are  updated starting from given initial conditions until a fixed point of the iteration if found.
At the fixed point the messages $y^{A(B)}_{i\to j}=\lim_{n\to \infty} y^{A(B),n}_{i\to j}$ satisfy the following relation
\begin{eqnarray}
y^{A}_{i\to j}&=&\left[1-\prod_{k\in N_A(i) \setminus j}(1-y^{A}_{k \to i})\right]\prod_{k\in N_B(i)} (1-y^{B}_{k\to i})\nonumber \\
y^{B}_{i\to j}&=&\left[1-\prod_{k\in N_B(i) \setminus j}(1-y^{B}_{k \to i})\right]\prod_{k\in N_A(i)} (1-y^{A}_{k\to i})
\label{ant_message}
\end{eqnarray}
If we average the Eqs $(\ref{ant_S})$ and $(\ref{ant_message})$ over an ensemble of networks with degree distribution $p^A(k),p^{B}(k)$ we get the equation for $S_{A(B)}=\avg{S^{A(B)}_i}$ and $S_{A,B}^{\prime}=\avg{y^{A(B)}_{k\to i}}$, where    $S_{A(B)}$  is the probability to find a node in the percolation cluster of network A(network  B), and $S_{A(B)}^{\prime}$ is the probability that following a link we reach a node in the percolation cluster of network A (network B). In particular, we have
\begin{eqnarray}
S_A&=&[1-G_0^A(1-S_A^{\prime})]G_0^B(1-S_B^{\prime})\nonumber\\
S_B&=&[1-G_0^B(1-S_B^{\prime})]G_0^A(1-S_A^{\prime}).
\label{g0int}
\end{eqnarray}
In Eq.~$(\ref{g0int})$ we have used $G_0^{A(B)}(z)$ and $G_1^{A(B)}(z)$ to indicate the generating functions of network A and B defined according to the definition
\begin{eqnarray}
G_1(z)&=&\sum_k \frac{k p_k}{\avg{k}}z^{k-1}\nonumber \\
G_0(z)&=&\sum_k p_k z^k,
\label{Gen}
\end{eqnarray}
where we use the degree distributions $p^A(k),p^B(k)$, respectively, for network A and network B.
Moreover $S_{A(B)}^{\prime}$  on a locally tree like network, satisfy the following recursive equations
\begin{eqnarray}
\hspace*{-3mm}S_A^{\prime}&=&(1-G_1^A(1-S^{\prime}_A))G_0^B(1-S_B^{\prime})=f_A(S_A^{\prime},S_B^{\prime}),\nonumber\\
\hspace*{-3mm}S_B^{\prime}&=&(1-G_1^B(1-S^{\prime}_B))G_0^A(1-S_A^{\prime})=f_B(S_A^{\prime},S_B^{\prime}).
\label{rec}
\end{eqnarray}

\begin{figure}
%\center
\includegraphics[width=0.75\textwidth]{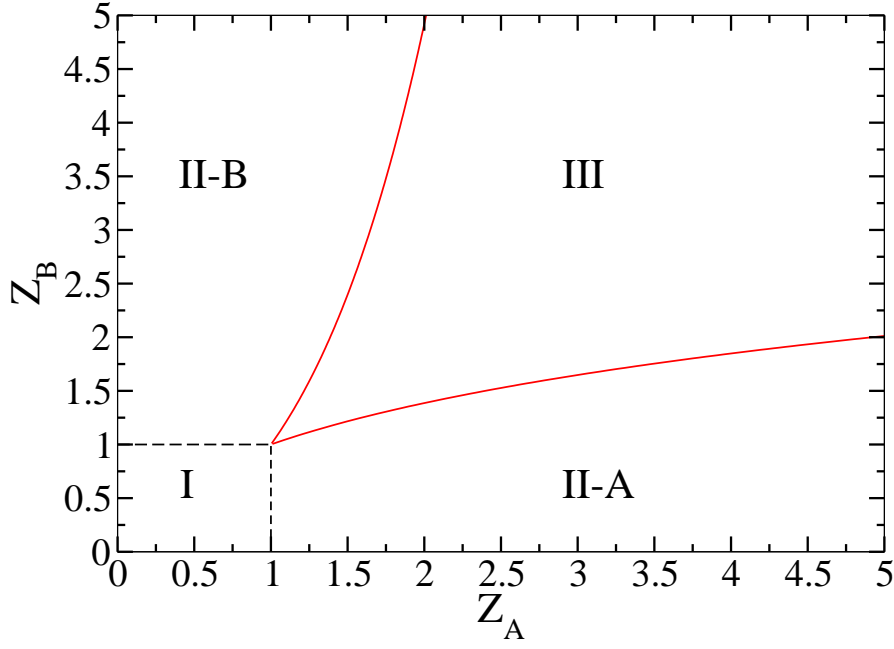}
\caption{Phase diagram of the percolation process on two antagonistic Poisson networks of average degree $\avg{k}_A=z_A$ and $\avg{k}_B=z_B$ respectively.}
\label{ER_ER}
\end{figure}

The solutions to the recursive Eqs.~$(\ref{rec})$  can be classified into three categories:
%\begin{itemize}
%\item

{\it (i) }The trivial solution in which neither of the network is percolating
$S^{\prime}_A=S^{\prime}_B=0$.
%\item

{\it {(ii)} }The solutions in which just one network is percolating.
In this case we  have either  $S^{\prime}_A>0,S^{\prime}_B=0$ or $S^{\prime}_B>0,S^{\prime}_A=0$.
From  Eqs.~$(\ref{rec})$ we find that the solution $S^{\prime}_A>0,S^{\prime}_B=0$ emerges  at a critical line of second order phase transition,  characterized by the condition
\begin{equation}
\left. \frac{d G_1^{A}(z)}{dz}\right|_{z=1}\equiv\frac{\avg{k(k-1)}_{A}}{\avg{k}_{A}} =1.
\end{equation}
Similarly the solution $S^{\prime}_B>0,S^{\prime}_A=0$ emerges at a second order phase transition when we have $\frac{\avg{k(k-1)}_{B}}{\avg{k}_{B}}=1$.
This condition is equivalent to the critical condition for percolation in single networks, as it should,  because one of the two networks is not percolating.
%\item

\begin{table}
\begin{indented}
\item[]\begin{tabular}{@{}ll}
\br
Region I   & $S_A^{\prime}=S_B^{\prime}=0$\\
Region II-A &$S_A^{\prime}>0, S_B^{\prime}=0$\\
Region II-B & $S_B^{\prime}>0,S_A^{\prime}=0$\\
Region III & either $S_A^{\prime}>0, S_B^{\prime}=0$ or $S_B^{\prime}>0,S_A^{\prime}=0$\\
\br
\end{tabular}
\end{indented}
\caption{Stable phases in the different regions of the phase diagram of  the percolation problem on two antagonistic Poisson networks (Figure $\ref{ER_ER}$).}
\label{t1}
\end{table}

{\it {(iii)} }The solutions  for which both networks are percolating. In this case we have $S^{\prime}_A>0,S^{\prime}_B>0$.
This solution can either emerge   (a) when the curves $S_A^{\prime}=f_A(S_A^{\prime},S_B^{\prime})$ and $S_B^{\prime}=f_B(S_A^{\prime},S_B^{\prime})$ cross at $S_A=0$ or at  $S_B=0$   (b) when the curves $S_A^{\prime}=f_A(S_A^{\prime},S_B^{\prime})$ and $S_B^{\prime}=f_B(S_A^{\prime},S_B^{\prime})$ cross at a point $S_A\neq 0$and $S_B\neq 0$ in which they are tangent to each other. For situation (a) the critical line can be determined by imposing, for example, $S_A^{\prime} \rightarrow 0$  in Eqs.~(\ref{rec}), which yields
\begin{eqnarray}
S^{\prime}_{B}&=&1-G_1^{B}(1-S^{\prime}_{B}), \nonumber \\
1&=& \frac{\avg{k(k-1)}_{A}}{\avg{k}_{A}} G_0^{B}(1-S^{\prime}_{B}).
\label{crit2}
\end{eqnarray}
A similar system of equation can be found by using Eqs.~$(\ref{rec})$ and imposing $S_B^{\prime}\rightarrow 0$.
For situation (b) the critical line can be determined imposing that the curves $S_A^{\prime}=f_A(S_A^{\prime},S_B^{\prime})$ and $S_B^{\prime}=f_B(S_A^{\prime},S_B^{\prime})$, are tangent to each other at the point where they intercept. This condition can be written as
\begin{equation}
\left(\frac{\partial{f_A}}{\partial S_A^{\prime}}-1\right) \left(\frac{\partial{f_B}}{\partial S_B^{\prime}}-1\right)-\frac{\partial{f_A}}{\partial S_B^{\prime}}\frac{\partial{f_B}}{\partial S_A^{\prime}}=0,
\label{tangent}
\end{equation}
where $S_A^{\prime},S_B^{\prime}$ must satisfy the Eqs.~(\ref{rec}).
%\end{itemize}

\subsection{The stability of the solutions}
Not every solution of the recursive Eqs.~$(\ref{rec})$ is stable.
Therefore here we check the  stability of the fixed points solutions of Eqs. $(\ref{rec})$  by linearizing the equations around each solution.  The  Jacobian matrix $J$  of the system of  Eqs. $(\ref{rec})$ is given by 
\begin{equation}
J=\left| \begin{array}{cc}
          \frac{\partial f_A}{\partial S_A^{\prime}} & \frac{\partial f_A}{\partial S_B^{\prime}} \\ 
	  \frac{\partial f_B}{\partial S_A^{\prime}} & \frac{\partial f_B}{\partial S_B^{\prime}} 
         \end{array}
   \right|.
\end{equation}
 The eigenvalues  $\lambda_{1,2}$ of the Jacobian can be found  by solving the characteristic equation $|J-\lambda I|=0$, which reads for our specific problem,
\begin{equation}
\left(\frac{\partial{f_A}}{\partial S_A^{\prime}}-\lambda \right) \left(\frac{\partial{f_B}}{\partial S_B^{\prime}}-\lambda \right)-\frac{\partial{f_A}}{\partial S_B^{\prime}}\frac{\partial{f_B}}{\partial S_A^{\prime}}=0.
\label{stability}
\end{equation}
The change of stability of each solution will occur when $\max(\lambda_1,\lambda_2)=1$.
 In the following we will discuss the stability of the solutions of type (i)-(iii).
 \begin{itemize}
 \item {\it (i) Stability of the trivial solution $S_A^{\prime}=S_B^{\prime}=0$.}
The solution is stable as long as the following two conditions are satisfied:
\begin{eqnarray}
\lambda_{1,2}=\frac{\avg{k(k-1)}_{A/B}}{\avg{k}_{A/B}} < 1.
\end{eqnarray}
Therefore the stability of this solution change on the critical lines $\frac{\avg{k(k-1)}_{A}}{\avg{k}_{A}}=1$ and $\frac{\avg{k(k-1)}_{B}}{\avg{k}_{B}}=1$.
\item{\it (ii) Stability of the solutions in which only one network is percolating.}
For the case of $S_A^{\prime}=0$ $S_B{\prime}>0$ the stability condition reads
\begin{eqnarray}
\lambda_1&=&\left.\frac{G_1(z)}{dz}\right|_{z=1-S^{\prime}_{B}}<1 \nonumber \\
\lambda_2&=&\frac{\avg{k(k-1)}_{A}}{\avg{k}_{A}} G_0^{B}(1-S^{\prime}_{B})<1.
\end{eqnarray}
We note here that if $\lambda_2>\lambda_1$ we expect to observe a change in the stability of the solution on the critical line given by Eqs. $(\ref{crit2})$.
A similar condition holds for the stability of the solution ${S_A^{\prime}>0,S_B^{\prime}=0}$.
\item{\it (iii) Stability of the solution in which  both networks are percolating $S_A^{\prime}>0,S_B^{\prime}>0$}
For characterizing  the stability of the solutions of type III we have to solve  Eq.(\ref{stability}) and impose that the eigenvalues $\lambda_{1,2}$ are less then 1, i.e. $\lambda_{1,2}<1$.
We observe here that for $\lambda=1$  Eq.(\ref{stability}) reduces to Eq.(\ref{tangent}). Therefore we expect to have a stability change of these solutions on the critical line given by Eq. $(\ref{tangent})$. 
\end{itemize}

\subsection{Two Poisson networks.}
In order to consider a specific example of antagonistic networks  we consider two antagonistic Poisson networks with average degree $\avg{k}_A=z_A$ and $\avg{k}_B=z_B$ respectively. 
In the  case of a Poisson network we have $G_0(z)=G_1(z)=e^{\avg{k}(1-z)}$. Therefore we have $S_A=S_A^{\prime}$ and $S_B=S_B^{\prime}$. The recursive equations Eqs.$(\ref{rec})$ read in this case
\begin{eqnarray}
S_A=(1-e^{-z_A S_A})e^{-z_B S_B}\nonumber \\
S_B=(1-e^{-z_B S_B})e^{-z_A S_A}.
\end{eqnarray}
In  Table $\ref{t1}$ we  characterize the phase diagram  percolation  on two antagonistic Poisson networks shown in Fig.~$\ref{ER_ER}$. The critical lines are given by $z_A=1$, $z_B=1$ and by $z_B=\frac{\log z_A}{1-1/z_A}$ or $z_A=\frac{\log(z_B)}{1-1/z_B}$. In particular we observe a first order phase transition along the line $z_A>1$, and  $z_B=\log(z_A)/(1-1/z_A)$ and along the line $z_B>1$ $z_A=\log(z_B)/(1-1/z_B)$ indicated as a solid red lines in Figure $\ref{ER_ER}$. The other lines indicated as black dashed lines in Figure $\ref{ER_ER}$ are critical lines of a second order transition.

One should note that the  solution $S_A^{\prime}>0,S_B^{\prime}>0$ in which both networks are percolating is always unstable in this case. This implies  that  for each realization of the percolation process, only  one of the two networks is percolating.

\begin{figure}
\center
\includegraphics[width=1.0\columnwidth ]{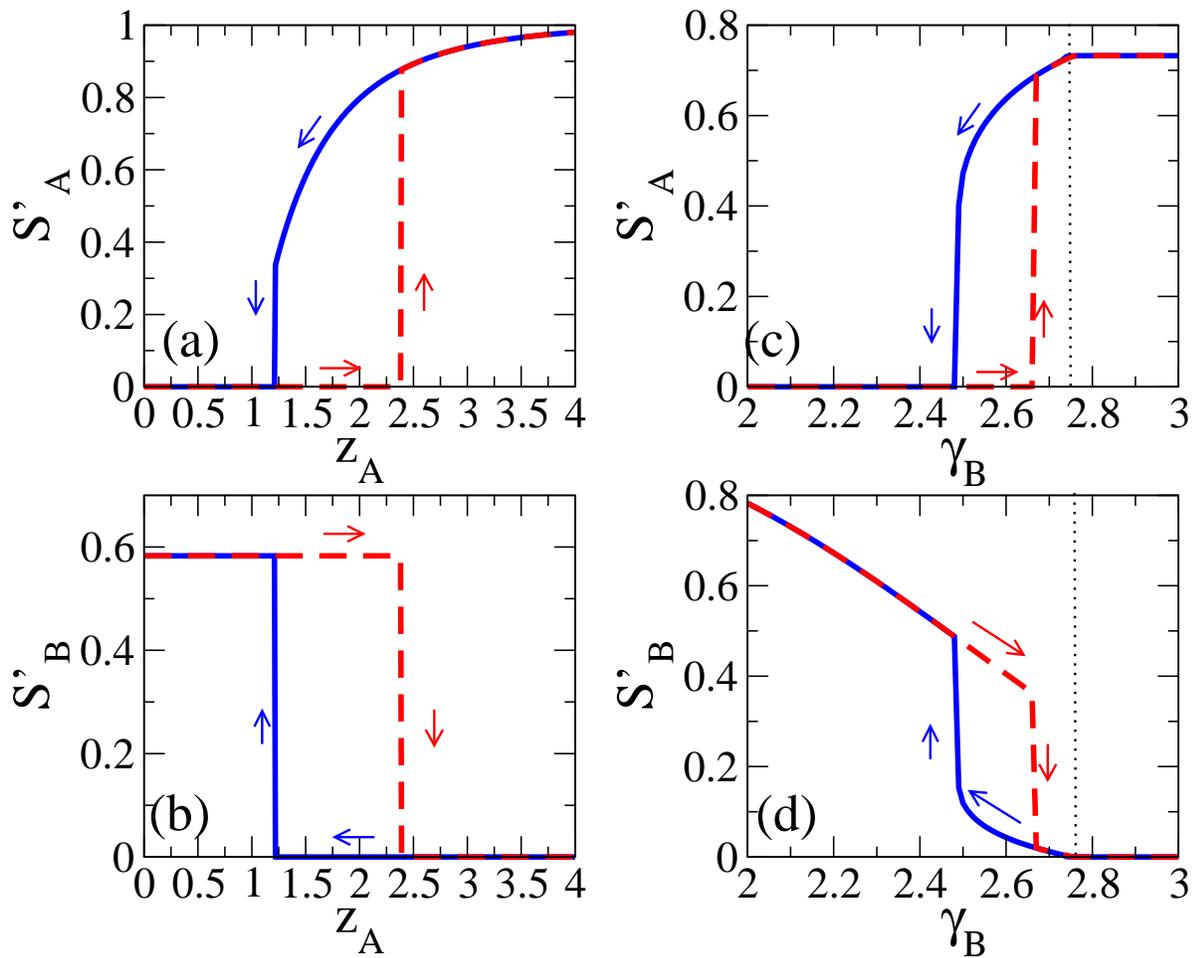}
\caption{(Color online) Panels (a) and (b) show the hysteresis loop for the percolation problem on two antagonistic Poisson networks with $z_B=1.5$. Panels (c) and (d) show the hysteresis loop for the percolation problem on two antagonistic networks of different topology: a Poisson network of average degree $z_A=1.8$ and a scale-free networks with power-law exponent $\gamma_B$, minimal degree $m=1$ and maximal degree $K=100$. The hysteresis loop is performed using the method explained in the main text.  The value of the parameter $\epsilon$ used in this figure  is $\epsilon=10^{-3}$.}
\label{hysteresis}
\end{figure}
In order to demonstrate the bistability of the percolation solution in region III  of the phase diagram we solved recursively the Eqs. $(\ref{rec})$ for $z_B=1.5$ and variable values of  $z_A$ (see Figure~\ref{hysteresis}). We start from values of $z_A=4$, and we  solve recursively the  Eqs.~$(\ref{rec})$. We find the solutions  $S_A^{\prime}=S_A^{\prime}(z_A=4)>0$, $S_B^{\prime}=S_B^{\prime}(z_A=4)=0$.Then we lower slightly $z_A$  and we solve again the Eqs.~$(\ref{rec})$ recursively, starting from the initial condition {$S_A^{\prime o}=S_A^{\prime}(z_A=4)+\epsilon$}, $S_B^{\prime o}=S_B^{\prime}(z_A=4)+\epsilon$, and  plot the result.  (The small perturbation $\epsilon>0$ is necessary in order not to end up with the trivial solution $S_A^{\prime}=0,S^{\prime}_B=0$.) Using this procedure we show that if we first lower the value of $z_A$  and then again we raise it, spanning the region III of the phase diagram as shown in Figure~$\ref{hysteresis}$ Panel (a) and (b), the solution present  an hysteresis loop. 
This means that in the region III either network A or network B might end up to be percolating.

\subsection{Two scale-free networks}
\begin{figure}
\center
\includegraphics[width=1.0\textwidth]{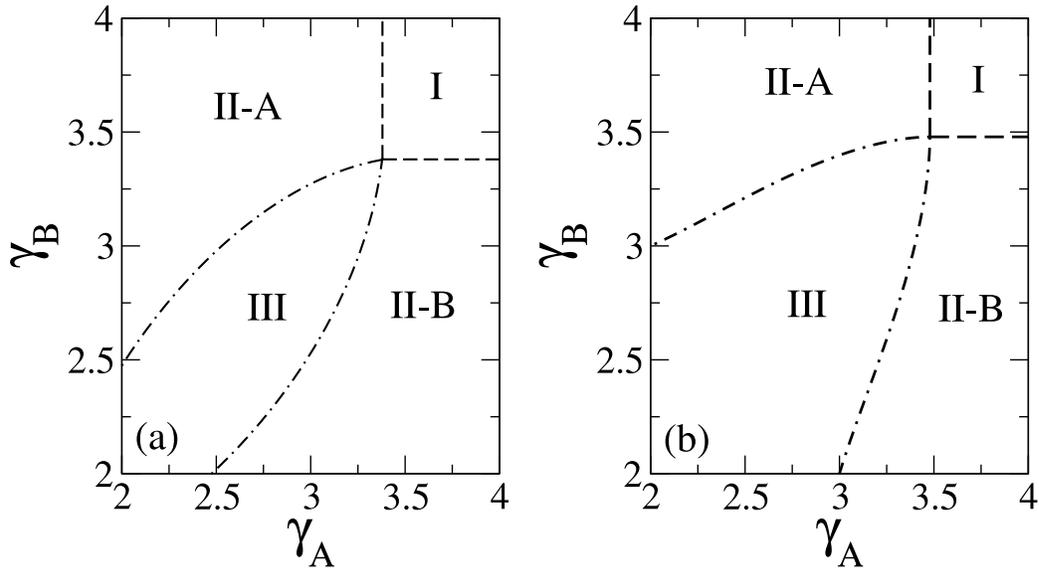}
\caption{The  phase diagram of the percolation process in two antagonistic  scale-free networks with power-law exponents $\gamma_A, \gamma_B$. The minimal degree of the two networks is $m=1$ and the maximal degree $K$. Panel (a) show the effective phase diagram with $K=100$, the panel (b) show the  phase diagram in the limit of an infinite network $K=\infty$.}
\label{SF_SF}
\end{figure}

Here, we characterize  the phase digram of two antagonistic scale-free  networks with power-law exponents $\gamma_A, \gamma_B$, as shown in Figure~$\ref{SF_SF}$. The two networks have minimal connectivity $m=1$ and varying value of the maximal degree $K$.

The critical lines of the phase diagram depend on the value of the maximal degree $K$ of the networks. The critical lines of the phase diagram are dependent on the value of the cutoff $K$ of the scale-free degree distribution and therefore for finite value of $K$ we observe an  effective phase diagram converging  in the $K\to \infty$ limit to the phase diagram of an infinite network.
In the infinite network limit the recursive equations Eqs. $(\ref{rec})$ can be written as 
\begin{eqnarray}
S_A^{\prime}=\left(1-\frac{\mbox{Li}_{\gamma_A-1}(1-S_A^{\prime})}{(1-S_A^{\prime})\zeta(\gamma_A-1)}\right)\frac{\mbox{Li}_{\gamma_B}(1-S_B^{\prime})}{\zeta(\gamma_B)}\nonumber \\
S_B^{\prime}=\left(1-\frac{\mbox{Li}_{\gamma_B-1}(1-S_B^{\prime})}{(1-S_B^{\prime})\zeta(\gamma_B-1)}\right)\frac{\mbox{Li}_{\gamma_A}(1-S_A^{\prime})}{\zeta(\gamma_A)},
\end{eqnarray}
where $\zeta(s)$ is the Riemann zeta function and Li$_n(z)$ is the polylogarithm function.
Solving these equations, and studying their stability as described in the previous paragraphs we can draw the phase diagram of the model.
The phase diagram is rich, showing a region (Region III) in the figure where both networks are percolating demonstrating an interesting interplay between the percolation  and the topology of the network.
The hub nodes of a network are the nodes which are more likely to be active in that network.Therefore hub nodes  play as a sort of "pinning centres" for the percolating cluster. Since the two antagonistic networks in our model have uncorrelated degrees, a node that is a hub in a network is unlikely to be a hub also in the other network offering the chance of having two percolating clusters in the two antagonist networks. This observation offers a qualitative understanding of why in two antagonist uncorrelated networks we can observe the coexistence of two percolating clusters while in the case of two Poisson networks where the degrees of the nodes are more homogeneous this phase is not observed.
The importance of the hub nodes in pinning the percolation cluster on one network can also help understand qualitatively the strong effects that a finite upper cutoff $K$ in the degree has  in the phase diagram.
A description of  the stable phases in the different regions of the phase diagram is provided by Table \ref{t2}. In this case all the transitions are second order.

\begin{table}
\begin{indented}
\item[]\begin{tabular}{@{}ll}
\br
Region I   & $S_A^{\prime}=S_B^{\prime}=0$\\
Region II-A &$S_A^{\prime}>0, S_B^{\prime}=0$\\
Region II-B & $S_B^{\prime}>0,S_A^{\prime}=0$\\
Region III &$S_A^{\prime}>0, S_B^{\prime}>0$ \\
\br
\end{tabular}
\end{indented}
\caption{Stable phases in the different regions of the phase diagram of  the percolation on two antagonistic scale-free networks (Figure $\ref{SF_SF}$).}
\label{t2}
\end{table}

\begin{figure}
\center
\includegraphics[width=1.0\textwidth]{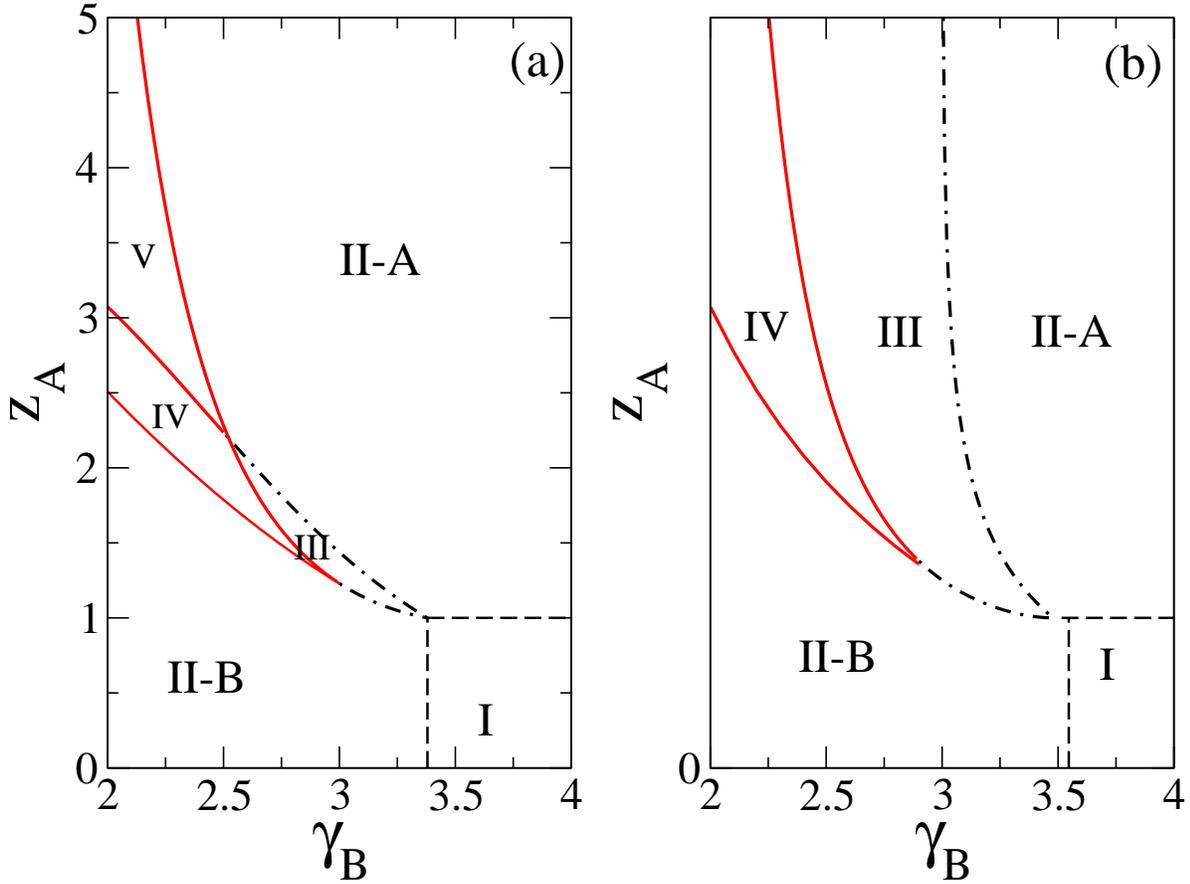}
\caption{(Color online) Phase diagram of the percolation process on a Poisson network with average degree $\avg{k}_A=z_A$ interacting with a scale-free network of power-law exponent $\gamma_B$, minimal degree $m=1$. Panel (a) show the effective phase diagram for maximal degree $K=100$, panel (b) show  phase diagram in the limit of an infinite network, i.e. $K=\infty$.}
\label{ER_SF}
\end{figure}

\begin{table}[h]
\begin{indented}
\item[]\begin{tabular}{@{}ll}
\br
Region I   & $S_A^{\prime}=S_B^{\prime}=0$\\
Region II-A &$S_A^{\prime}>0, S_B^{\prime}=0$\\
Region II-B & $S_B^{\prime}>0,S_A^{\prime}=0$\\
Region III & $S_A^{\prime}>0, S_B^{\prime}>0$\\
Region IV & either $S_B^{\prime}>0, S_A^{\prime}=0$ or  $S_A^{\prime}>0,S_B^{\prime}>0$\\
Region V & either $S_A^{\prime}>0, S_B^{\prime}=0$ or  $S_B^{\prime}>0,S_A^{\prime}=0$\\
\br
\end{tabular}
\end{indented}
\caption{Stable phases in the phase diagram for the percolation on two antagonistic networks: a Poisson network (network A) and a scale-free network (network B). (Figure $\ref{ER_SF}$)}
\label{t3}
\end{table}

\subsection{A Poisson network and a scale-free network}
Finally we  consider the case of a Poisson  network (network A) with average connectivity $\avg{k}_A=z_A$, and a network B with scale-free degree distribution and  power-law exponent of the degree distribution $\gamma_B$. The scale-free network has minimal connectivity $m=1$ and maximal degree given by  $K$.
In Figure $\ref{ER_SF}$ we show the phase diagram of the model in the plane $(\gamma_B,z_A)$.
The critical lines of the phase diagram are dependent on the value of the cutoff $K$ of the scale-degree distribution and therefore for finite value of $K$ we observe an  effective phase diagram converging  in the $K\to \infty$ limit to the phase diagram of an infinite network. In the infinite network limit the recursive equations Eqs. $(\ref{rec})$ can be written as 
\begin{eqnarray}
S_A^{\prime}=\left(1-e^{-z_AS_A}\right)\frac{\mbox{Li}_{\gamma_B}(1-S_B^{\prime})}{\zeta(\gamma_B)}\nonumber \\
S_B^{\prime}=\left(1-\frac{\mbox{Li}_{\gamma_B-1}(1-S_B^{\prime})}{(1-S_B^{\prime})\zeta(\gamma_B-1)}\right)e^{-z_A S_A}
\end{eqnarray}
where $\zeta(s)$ is the Riemann zeta function and Li$_n(z)$ is the polylogarithm function.
Solving these equations and studying their stability as described in the previous paragraphs we could draw the phase diagram of the model.
The phase diagram includes two  regions, (region III and region V) with bistability of the solutions and two regions (region III,  and region IV) in which the solution in which both networks are percolating is stable. 
We have indicated with red solid lines the lines where a first order phase transition can be observed in correspondence of a change of the stability of the solutions and we have indicated with black dashed lines the lines of a second order phase transition.
Also in this case the importance of hubs as centres for the pinning of a percolating phase is apparent. In fact we observe strong effects of a finite cutoff $K$ in the degrees of the nodes leading to the 
 disappearance of region V  when the cutoff $K$ goes to infinity.
 In fact the reason why  region V is only observed in the finite size network can be explained  with the effect that highly connected hubs have in the stabilization of a percolation phase in  network B.
The region V in fact contains a phase in which network B is not percolating, this phase is allowed only if the hubs are below a certain connectivity. Therefore this phase disappears in the limit of an infinite network.
In Table $\ref{t3}$ we describe the percolation stable solutions in the different regions of the phase diagram shown in Figure~$\ref{ER_SF}$.

In order to demonstrate the bistability of the percolation problem we solved recursively the Eqs. $(\ref{rec})$ for $z_B=1.8$ (see Figure~$\ref{hysteresis}$). We start from values of $\gamma_B=3$, and we  solve the  Eqs. $(\ref{rec})$ using the same method explained for the two antagonistic Poisson networks.
Using this procedure we show  in Figure $\ref{hysteresis}$ Panels (c) and (d) that  the solution present  a second order phase transition to a phase in which both networks are percolating and also an hysteresis loop in correspondence of region IV. This  demonstrates the bistability of the solutions in region IV and the existence of a phase in which both network percolate in region III.

\section{Conclusions}
In conclusion,  we have investigated  how much antagonistic interactions modify the phase diagram of the percolation transition.
The percolation process on two antagonistic networks shows important new physics of the percolation problem. In fact, the percolation process in this case shows a bistability of the solutions. This implies   that the steady state of the system is not unique.
In particular, we  have demonstrated the bistability of the percolation solution for the percolation problem on two antagonistic Poisson networks, or two antagonistic networks with different topology: a Poisson network and a scale-free network.
Moreover, in the  percolation transition between two scale-free antagonistic networks and in the percolation transition between two antagonistic networks with  a Poisson network and a scale-free networks, we found a region in the phase diagram in which both networks are percolating, despite the presence of antagonistic interactions.
We believe that this paper opens new perspectives in the percolation problem on interdependent networks,
which might include both interdependencies and antagonistic interactions eventually combined in a boolean rule.
In an increasingly interconnected world, understanding how much these different types of interactions affect percolation transition is becoming key to answering fundamental questions about  the robustness of interdependent networks.

\end{document}